\documentclass[pra,aps,twocolumn,floatfix,citeautoscript,showpacs]{revtex4-2}
\usepackage{graphicx,rotating,subfigure,amsmath,amsfonts,amssymb,delarray,color,subcaption}
\usepackage{caption} 
\usepackage{hyperref}
\usepackage{xcolor}
\usepackage{soul}
\usepackage{dsfont}
\usepackage[T1]{fontenc}
\usepackage{physics}
\usepackage{multirow}
\usepackage{comment}
\newcommand{\im}{\text{i}}

\def\12{\frac{1}{2}}

\begin{document}
	\title{Limits of the non-Hermitian description of decay models}
	\author{Kyle Monkman$^1$}
	\author{Mona Berciu$^{1,2}$}
	\affiliation{$^1$Quantum Matter Institute, University of British Columbia, Vancouver, British Columbia V6T 1Z4, Canada}
	\affiliation{$^2$Department of Physics and Astronomy, University of British Columbia, Vancouver, British Columbia, Canada,V6T 1Z1}
	\date{\today}
	\begin{abstract} 
		We present a general proof that  non-Hermitian dynamics and Lindblad dynamics with only decay terms are equivalent in the highest particle subspace. We then propose an unbiased method to determine if a system's dynamics in the highest-particle subspace is non-Hermitian. We exemplify this for a simple two-site decay system connected to two baths, and find that the exact solution is well approximated by non-Hermitian dynamics only in the weak-coupling and in the singular-coupling limits, where a Lindbladian description was already known to be accurate. The fact that an accurate non-Hermitian description is so limited, even for such a simple system, raises doubts about how valid such descriptions are for more complicated systems away from these asymptotic limits. Finally, we prove that for models with a non-degenerate system Hamiltonian,  exceptional points cannot occur in the weak coupling limit. This result is relevant for the design of experiments that aim to identify such exceptional points.
	\end{abstract}
	\maketitle 
	\section{Introduction} A great challenge of modern physics is to identify and understand the novel quantum phenomena that occur in open systems \cite{BreuerPetruccione, RaimondHaroche}. This requires a good understanding of the theoretical frameworks that can be used to accurately describe experimental data \cite{Murch, Murch2, Wineland, Zhang, IBM}, which, in turn, motivates the
	important goal of  theoretical physics to study how to properly `integrate out' the infinite number of degrees of freedom of a bath model. 
	
	One extensively used approach for modelling open systems is the Lindblad master equation \cite{Lindblad1976, GKS, manzano2020short}, which replaces solving the Schr\"odinger equation for the total Hamiltonian
	\begin{equation}
		\label{Open}
		H=H_A+H_B+H_C
	\end{equation}
	where $H_A$,  $H_B$  and $H_C$ describe the system, the bath(s), and  their coupling, respectively,  with the Lindblad equation for the density matrix of the system 
	\begin{multline}
		\label{Lindblad}
		\dot{\rho}_A=\mathcal{L}[\rho_A]= -\im [H_A,\rho_A] \\
		+ \sum_j \Gamma_j \left( L_j \rho_A L_j^\dag -\frac{1}{2} \left\{ L_j^\dag L_j , \rho_A \right\} \right).
	\end{multline}
	Here $L_j$ are system operators, the gain/decay rates  $\Gamma_j$ vanish when $H_C=0$, and we have set $\hbar=1$.
	
	The Lindblad master equation describes an open quantum system as being connected to a Markovian (memoryless) bath. This Markovian description of a bath has been shown to be accurate (i) in the weak coupling limit, where the energy scale $C$ of the system-bath coupling is much smaller than the  energy scales $A, B$ of the system and of the bath; and (ii) in the singular limit where $B/A\rightarrow \infty$ and $C/A\rightarrow \infty$ such that $C^2/B$ is constant \cite{BreuerPetruccione,palmer1977singular}. Given its relative simplicity, the Lindblad equation has been used extensively to interpret experimental results \cite{Murch, Murch2, Wineland, Zhang, IBM}, regardless of whether or not the system demonstrably falls in the limits (i) or (ii).

	Another very popular approach for the study of open systems is to assume that the dynamical equation for $\rho_A$ is of the form $\dot{\rho}_A= -\im [H_{nh},\rho_A]$, for some non-Hermitian operator $H_{nh}$ \cite{Sato, Bardarson, Nunnenkamp, Lorenzo,Porras, Monkman, Budich, Purkayastha, Larson, Srivatsa, Barch, MardaniSirker,Wang,Wu}. However, it is still far from clear when such a non-Hermitian description of an open system is adequate.
	
	A direct link between the two approaches is  established by noting that if the action of the quantum jumps term
	\begin{equation}
		\label{LindbladQuantumJumps}
		\mathcal{Q}[\rho_A]=\sum_j \Gamma_j L_j \rho_A L_j^\dag
	\end{equation}
	can be ignored in Eq. (\ref{Lindblad}), then the Lindblad equation is equivalent to evolution under the non-Hermitian operator \cite{Nori, Bagarello}
	\begin{equation}
		\label{LindbladHamiltonian}
		H_{\text{nh}}=H_A -\frac{\im}{2}\sum_j \Gamma_j {L}^\dag_j L_j.
	\end{equation}
	Furthermore, because the right eigenvectors of this non-Hermitian operator are not necessarily orthogonal to one another, this opens the possibility to observe exceptional points when two eigenvectors of $H_{\text{nh}}$ become parallel to one another \cite{Flore, Nori, Bagarello, Murch,hatano2019exceptional}. 
	
	In this article, we investigate some of the questions regarding the existence of accurate non-Hermitian and Lindbladian descriptions for open systems, and offer some answers for the case where the baths are {\em completely empty} at the initial moment. 
	
	The first result of this article (R1) is to prove that for Lindblad dynamics {\em with only decay terms} (see below for a formal definition), the action of the quantum jumps vanishes in the subspace with the highest particle number.  This provides an exact equivalence between such decay Lindblad dynamics and non-Hermitian dynamics in this subspace, through Eq. \eqref{LindbladHamiltonian}. As we detail below, if there are eigenstates of a non-Hermitian matrix in this subspace, they do not mix with other states within the subspace. This gives us a direct way to verify whether such eigenstates exist, and thus to check whether the dynamics within this subspace is defined by a non-Hermitian operator, or not.

\begin{figure}[t]
		\includegraphics[width=0.3\linewidth]{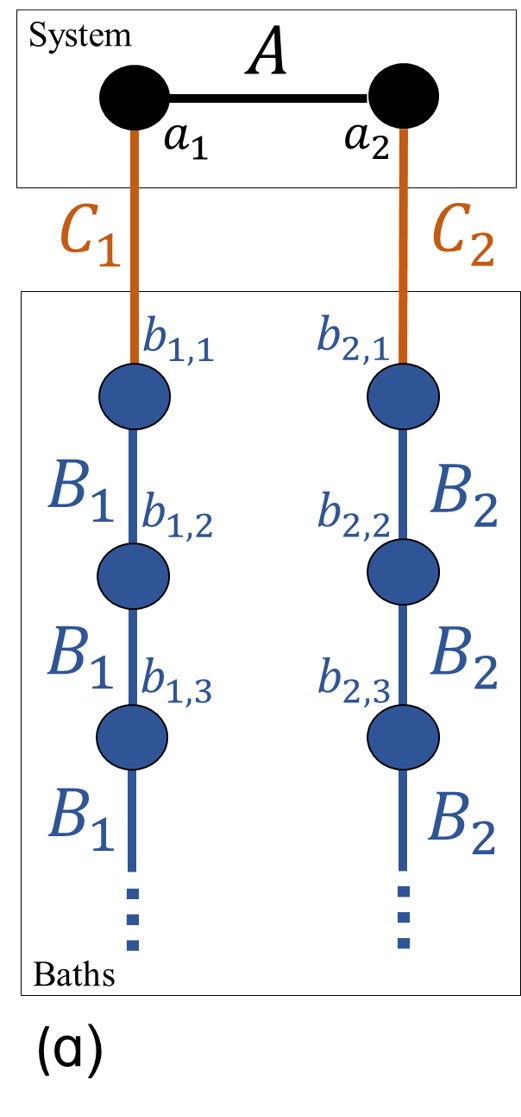} 
		\raisebox{0.2\height}{\includegraphics[width=0.6\linewidth]{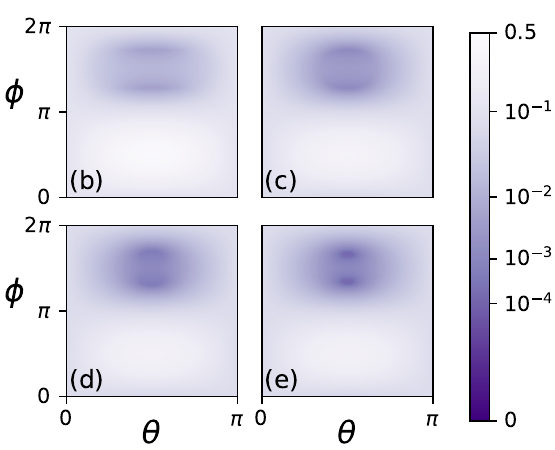}}  
		\caption{($a$) Sketch of the system+bath model $H=H_A+H_B+H_C$ described in eqs. \eqref{HA}, \eqref{HB} and \eqref{HC}. \color{black}($b$)-($e$) $R_{\theta,\phi}$ of Eq. \eqref{R}  as a function of $\theta$ and $\phi$ for $C_1=C$, $C_2=3C$, $B_1=B$, $B_2=2B$ and $\frac{C^2}{B}=0.5$, and  ($b$) $B=2$, $C=1$, ($c$) $B=4$, $C=\sqrt{2}$, ($d$) $B=8$, $C=2$ and ($e$) $B=16$, $C=2\sqrt{2}$. As the singular limit is approached, the two  minima gradually approach zero, indicating the appearance of two non-mixing states and validating non-Hermitian dynamics in this limit. 
			\label{fig:1abdce}}
	\end{figure}

	We investigate the existence of such eigenstates for a simple two-site system connected to two microscopically specified baths, see sketch in Figure \ref{fig:1abdce}a. We time evolve a single particle starting initially in system A and which decays out into one of the baths, eventually. We find that eigenstates exist in this single-particle subspace of system A \textit{only} (i) in the weak, and (ii) in the singular coupling limits, showing that a non-Hermitian description is only valid in these two limits. For brevity, we will refer to this result as R2. In our opinion, the fact that even for such a simple open system, an accurate non-Hermitian description is found only in those two narrow asymptotic regimes, raises doubts about how appropriate such non-Hermitian descriptions are in general.

	Furthermore, as mentioned above, it is already known that a Lindblad description must be valid in these two limits. Together with R1, this shows that a Lindbladian with only decay terms provides a good description of this open system in these two limits, but nowhere else in the parameter space.

	Our third  result (R3) is to show that for this class of bath models
	and for a non-degenerate $H_A$, exceptional points {\em cannot} occur in the weak coupling limit.
	They can occur in other regions where non-Hermitian dynamics is valid; for instance, for the simple model of Fig.  \ref{fig:1abdce}, exceptional points in the highest particle subspace can occur in the singular coupling limit. Physically, this means that an exceptional point can appear  when the intrinsic timescale of the isolated system is comparable to the relaxation time of the system. This result  will help guide experimental searches for open systems with exceptional points. 
	
	The rest of the article is organized as follows: Section \ref{Methodology} describes the general methodology we deploy. Section \ref{ModelAndResults} introduces the specific model we study, and the corresponding results, while Section \ref{Conclusions} has our conclusions. Many of the more mathematical details have been relegated into various Appendixes.

	\section{Methodology} 
	\label{Methodology}
	In this work, we only consider  quantum systems which initially have $N$ particles, and which are connected   to one or more {\it empty  baths} at time $t=0$. If the bath modeling is appropriate for the system, {\it i.e.} all the eigenstates of the isolated system are well within the baths' spectra so that particles can move away from the system into the baths, then the number of particles in the main subsystem vanishes as $t\rightarrow \infty$. We call such an open system, and any other system that can be mapped onto it, {\em a decay model}.

	In this description of decay, the total particle number $N$ of the system+baths is conserved. Therefore, we can always write the full system+baths state as $|\Psi \rangle=\sum_n x_n |\psi_n\rangle$ where $|\psi_n \rangle$ is a state with $n$ particles in the main system and $N-n$ particles in the baths. Then the reduced density matrix of the main system is in block diagonal form $\rho_A=\Tr_B[|\Psi\rangle\langle\Psi|]=\rho^{(0)} \oplus \rho^{(1)} \oplus \dots \oplus \rho^{(N)}$ where $\rho^{(n)}$ represents having $n$-particles in the system. We also assume that the system Hamiltonian $H_A$ conserves particle number in the main system. Therefore, it is also of block diagonal form $H_A=H^{(0)} \oplus H^{(1)} \oplus H^{(2)} \oplus \dots \oplus H^{(N)}$,
	where each block $H^{(n)}$ acts on the $n$-particle Hilbert space. 
	
	This type of situation can be described by a Lindblad master equation with decay terms. We define a {\em Lindblad evolution with only decay terms} by the property that all Lindblad operators $L_j$ map an $n$ particle state to an $n-1$ particle  state, {\em  i.e.} they are all annihilation operators. Then the evolution of the system density matrix in the highest particle Hilbert space, $\rho^{(N)}(t)$,  is non-Hermitian:
	\begin{equation}
		\label{Highest}
		\rho^{(N)}(t)=\exp(-\im H_{\text{nh}} t) \rho^{(N)}(0) \exp(\im H_{\text{nh}}^\dag t) 
	\end{equation}
	where $H_{nh}$ is given by Eq. \ref{LindbladHamiltonian}. This is R1, and holds because the quantum jumps have no effect within this particular subspace with the highest particle number. We fully prove  Eq. \eqref{Highest}  in Appendix \ref{R1appendix}. 
	
	We emphasize that one  can measure this dynamics with post-selective measurements, or with observables that only count the highest particle subspace. This is done in Ref. \cite{Murch, Murch2}, where the quantum jump terms only decay the particles outside of a designated Hilbert space.  Of course, the quantum jumps are essential for evolving $\rho_A$ into subspaces with fewer than $N$ particles, so the full evolution of $\rho_A(t)$ need not be non-Hermitian.

	Next, let us assume that for a decay model, there is a non-Hermitian $H_{nh}$ such that $\rho^{(N)}(t)$ is described by Eq. (\ref{Highest}), and let us discuss how the eigenstates of $H_{nh}$ can be detected  in this highest-particle subspace. Specifically, let $|v \rangle  $ be a right eigenvector of $H_{nh}$. If we choose $|v\rangle$ as the initial state then $\rho^{(N)}(t) \propto |v\rangle\langle v|$ (with a decaying pre-factor) for all times, {\it i.e.} this initial state does not mix into other  $N$-particle states. Explicitly, if $P_v=1-|v\rangle\langle v|$ is the projector onto  $N$-particle states orthogonal to $|v \rangle$, then
	\begin{equation}
		\label{r}
		r_v(t)= \Tr[\rho^{(N)}(t)P_v]=0
	\end{equation}
	at all times.
	
	For completeness,  it is important to note that the reciprocal is not necessarily true, {\it  i.e.} there can exist  vectors for which  $r_v(t)=0$ when  $\rho^{(N)}(t)$ is not described by non-Hermitian dynamics. This can happen  when there are additional global symmetries of the system+bath total Hamiltonian,  which protect subspaces invariant to these symmetries from mixing together (although such symmetries would not prevent mixing within each such subspace). We assume that such symmetries would be broken in a realistic system with any degree of imperfections, whereas if  $r_v(t)=0$  arises from non-Hermitian dynamics, it should be  robust to imperfections that remove global symmetries.

	To summarize, we can find possible right eigenvectors of $H_{nh}$ --and thus, test whether the evolution of $\rho^{(N)}(t)$ is  non-Hermitian (or not) for a system without global symmetries-- by scanning through all states $|v\rangle$ in the $N$-particle subspace, and finding all non-mixing states satisfying Eq.  \eqref{r}. Next, we use the exact solution of simple decay models to show that for such systems, the evolution of $\rho^{(N)}(t)$ is well approximated by  non-Hermitian dynamics only in the  weak- and singular-coupling limits, demonstrating R2.

	\section{Model and results}
	\label{ModelAndResults}
	Generally, decay models can be evolved exactly  from any initial state with any value of $N$, although the computational costs quickly become  prohibitive.  Here we consider an example with $N=1$,  which suffices because the Hamiltonian we consider is non-interacting. In what follows, we verify that for this simple model, such non-mixing vectors appear only in the weak- and singular-coupling limits, and therefore its dynamics is non-Hermitian only in these limits.

	The  specific model we study $H=H_A+H_B+H_C$ is sketched in Fig. \ref{fig:1abdce}a (its generalization to more sites and more baths is straightforward, see appendices \ref{GreensAppendix} and \ref{R3Appendix}).  It consists of a two-site system 
	\begin{eqnarray}
		\label{HA}
		H_A = A(a_1^\dag a_2 + a_2^\dag a_1)
	\end{eqnarray}
	where $a_s^\dag|0\rangle$ has a particle at site $s=1,2$. Each site is coupled to its own bath, described by  semi-infinite chains
	\begin{eqnarray}
		\label{HB}
		H_B = \sum_{s=1}^2 B_s \sum_{j=1}^\infty \left( b_{s,j}^\dagger b_{s,j+1}+b_{s,j+1}^\dag b_{s,j}  \right)
	\end{eqnarray}
	that allow particles to move inside the respective $s=1,2$ bath \cite{zhu2018excitonic, Plenio}. Finally, the coupling is  
	\begin{eqnarray}
		\label{HC}
		H_C = \sum_{s=1}^2 C_s \left( a_s^\dag b_{s,1} + b_{s,1}^\dag a_s \right).
	\end{eqnarray} 
	
	We use standard Green's functions methods \cite{zhu2018excitonic} (details in Appendix \ref{GreensAppendix})
	to calculate the time evolution of all  $N=1$ initial states parameterized as $|v \rangle= |\theta,\phi\rangle = (\cos \theta a_1^\dag+ e^{\im \phi} \sin\theta a_2^\dag)|0\rangle$. Because this $N=1$ subspace has dimension $d=2$, the projector  $P_v=1-|v\rangle\langle v|=|{\bar \theta},{\bar \phi}\rangle \langle {\bar \theta},{\bar \phi}|$, where $|{\bar \theta},{\bar \phi}\rangle = ( -e^{-\im \phi} \sin\theta a_1^\dag+ \cos \theta  a_2^\dag)|0\rangle $ is the 1-particle state orthonormal to $|v\rangle$. Thus, for any set values of $A,B_s,C_s$, we can scan the $\theta,\phi$ space and calculate
	\begin{equation}
		\label{R}
		R_{\theta,\phi}=\max_{t} |r_{v}(t)|
	\end{equation}
	over a finite time interval. $R_{\theta,\phi}$ must be equal to zero if the initial state $|v\rangle=|\theta, \phi\rangle$ is a non-mixing state.
	
	The symmetric case with $C_1=C_2$ and $B_1=B_2$ has a global inversion symmetry, and thus the states $|v_{\pm}\rangle =\frac{1}{\sqrt{2}}(a_1^\dag\pm a_2^\dag)|0\rangle$ do not mix, resulting in  $R_{\theta,\phi}=0$ for $\theta=\pi/2$ and $\phi=0,\pi$ strictly because of symmetry, regardless of the nature of the underlying dynamics. As mentioned, we ignore such a 'perfect' model, which is unlikely to be realized in practice.
	
	Instead, we consider a non-symmetric case with $C_2=3C_1=3C$, and $B_2=2B_1=2B$ (similar results are found for any other non-unity ratios). Next, we ask for which parameters $H_B$ acts as a valid bath for the system A, by which we mean that any particle initially in A can move freely into the bath. For our choices, the spectrum of $H_B$ ranges over  $[-4B,4B]$ while the spectrum of $H_A$  has eigenergies at $-A$ and $A$. So, intuitively, if $4B<A$, then a particle could remain trapped at all times in the subsystem $A$,  {\em i.e.} this is not an appropriate bath model. This intuition is correct, but the more precise valid-baths condition involves both $b_A=B/A$ and $c_A=C/A$ and requires that:
	\begin{equation}
		\label{bound}
		c_A^2<\frac{2b_A^2(x_0+y_0) - \sqrt{4b_A^4(x_0-y_0)^2+x_0 y_0 b_A^2}}{x_0 y_0}
	\end{equation}
	where $x_0=1/(2+\sqrt{3})$ and $y_0=9/2$ (see  Appendix \ref{GreensAppendix} for the derivation). Mathematically, this condition ensures that the single-particle Green's function which describes the dynamics of the main subsystem A does not have any poles on the real axis. 
	
	We calculate $R_{\theta,\phi}$ over the time interval $0 \le t \le 4\pi/A$ within the decay region, i.e. for parameters satisfying Eq. (\ref{bound}). In Figure \ref{fig:1abdce}, we set $\frac{C^2}{B}=0.5$ with ($b$) $B=2$, $C=1$, ($c$) $B=4$, $C=\sqrt{2}$, ($d$) $B=8$, $C=2$ and ($e$) $B=16$, $C=2\sqrt{2}$,  moving towards the singular coupling limit. The global minima in the subfigures are roughly equal to  ($b$) $0.0048$, ($c$) $0.0013$, ($d$) $0.0005$ and ($e$) $0.0001$. This demonstrates the evolution towards two non-mixing states (defined by $R_{\theta,\phi}=0$) in the singular coupling limit, which is approached asymptotically from panel ($b$) to ($e$).

	\begin{figure}[t]
		\includegraphics[width=0.99\linewidth]{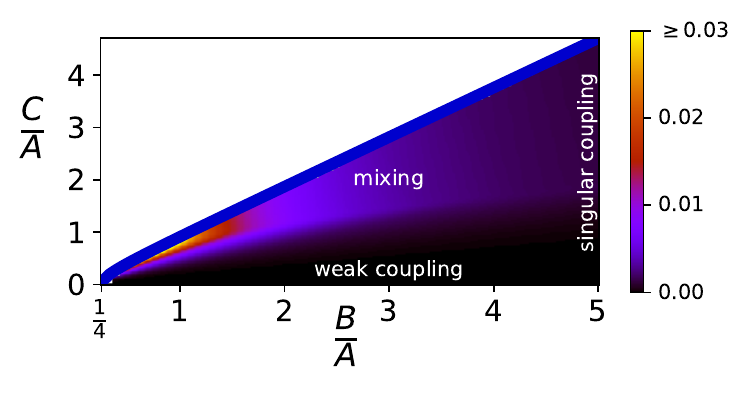} \includegraphics[width=0.85\linewidth]{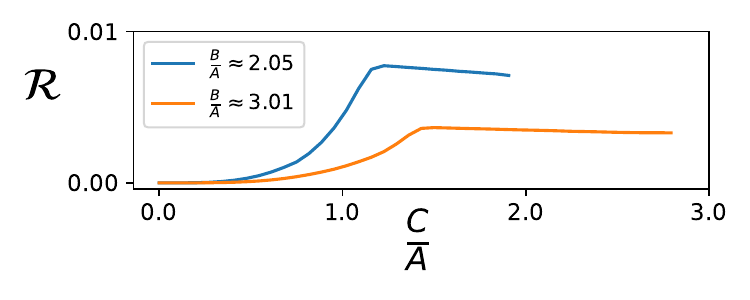}  \hfill \hfill \hfill
		\caption{Mixing parameter $\mathcal{R}$ of Eq. \eqref{RR} when  $C_1=C$, $C_2=3C$, $B_1=B$, $B_2=2B$.  Upper panel: colormap of $\mathcal{R}$ as a function of $\frac{C}{A}$ and $\frac{B}{A}$. The thick blue line is Eq. \eqref{bound} and the region of interest (where the baths are indeed  valid baths) is below it. Here, possible non-Hermitian dynamics is signalled by $\mathcal{R} \rightarrow 0$ (dark regions) and is approached asymptotically only in  the weak coupling region (bottom region) and in the singular coupling region (right side). Lower panel: $\mathcal{R}$ as a function of $\frac{C}{A}$ for $\frac{B}{A} \approx 2.05$ and $\frac{B}{A} \approx 3.01$, showing a finite $\mathcal{R}$ everywhere along those cuts, except in the weak-coupling limit.
			\label{fig:main}}
	\end{figure}

	To search for such  minima systematically, we use the global parameter
	\begin{equation}
		\label{RR}
		\mathcal{R}=\min_{\theta,\phi} R_{\theta,\phi}
	\end{equation}
	to characterize whether the evolution of $\rho_N(t)$ is (when $ \mathcal{R}=0$) or is not (when $ \mathcal{R}>0$) non-Hermitian. The considered time interval $0 \le t \le 4\pi/A$ is sufficient to verify which parameter regions of Figure \ref{fig:1abdce} definitely have mixing, $\mathcal{R}>0$. However, in and by itself, it does not exclude the possibility that either the weak and/or the singular limits might show mixing at some later $t>4 \pi/A$. Nevertheless, we believe this to be unlikely: The weak coupling limit  definitely has $\mathcal{R}=0$ when $C=0$ as the dynamics are exactly Hermitian in this case. Also, for this system, the solution in the singular coupling limit is provably equivalent to Lindblad dynamics with only decay terms (see Appendix \ref{R3Appendix} for a proof).

	Figure \ref{fig:main} illustrates such an analysis for the same system, i.e. $C_2=3C_1=3C$, and $B_2=2B_1=2B$ (similar results are found for any other system without global symmetries). The blue line separates the regions where the valid bath conditions of Eq. (\ref{bound}) are and are not obeyed. That is, the white region has poles in the relevant Green's function. The value of $\mathcal{R}$ is shown as a contour plot for the decay region in the  $(B/A, C/A)$ parameter space. The plot reveals that $ \mathcal{R}\rightarrow 0$ in  the bottom region of the plot (the weak coupling limit, $C/A\ll 1$) and in
	the asymptotic right-side of the plot (the singular coupling limit, $C/A\rightarrow \infty$, $B/A\rightarrow \infty$ with $\gamma=C^2/B={\rm const.}$). These results are consistent with non-Hermitian dynamics and therefore also Lindblad dynamics with only decay terms occurring only in the weak and singular limits. 
	
	\begin{figure}[t]
		\centering
		\includegraphics[width=0.935\linewidth]{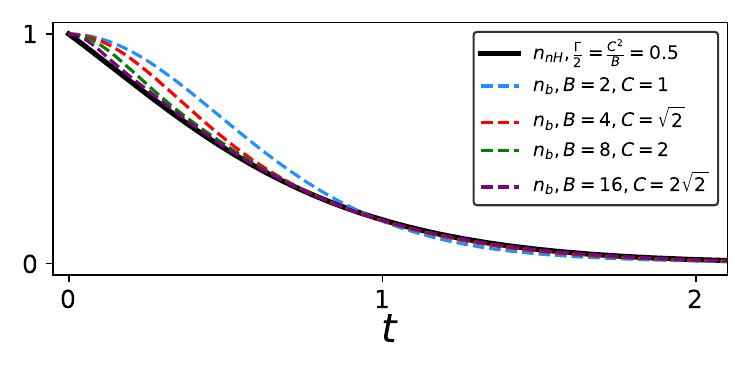} \hfill
		\caption{Comparison between $n_{nH}(t)$ (black line) and $n_b(t)$ (dashed lines) as a function of $t$. The accuracy of the non-Hermitian approximation improves as  the singular limit is approached. Here $A=1$.}
		\label{fig2}
	\end{figure}
	
	$\mathcal{R}$ is finite everywhere else  apart from  these two asymptotic limits, showing that the evolution of $\rho_{N=1}(t)$ here {\em cannot be non-Hermitian}, and thus
	proving that the evolution of $\rho_A(t)$ {\it is not captured by a Lindbladian {with only decay terms} anywhere else in the parameter space}. 
	
	Now we move our attention to our claim R3: if the spectrum of $H_A$ is non-degenerate, exceptional points cannot occur in the weak coupling limit. One can start to understand this based on physical grounds: if the couplings to the baths are perturbationally small, to zeroth order the eigenvectors of $H_{nh}$ are equal to the eigenvectors of $H_A$. The latter are always orthogonal and therefore distinct, and cannot be made identical by the addition of small corrections. In  Appendix \ref{R3Appendix} we use a perturbative argument \cite{kato2013perturbation} to prove this statement  for {\em any} system with a non-degenerate spectrum of $H_A$, and which has weak coupling to one or more baths like those considered in this article.

	Exceptional points may appear in other regions where the dynamics is non-Hermitian.
	As an example, we show that the model in Figure \ref{fig:1abdce}a can host exceptional points in the singular coupling limit. 
	
	For our simple model, the $N=1$ subspace dynamics in the singular limit  is described by $H_{nh}$ of Eq. (\ref{LindbladHamiltonian})
	with $\Gamma_s=2\frac{C_s^2}{B_s}$ and $L_s=a_s$. We verify this below numerically, and provide a mathematical proof in Appendix \ref{R3Appendix}.

	To demonstrate this claim, we calculate numerically the exact time-evolution of the expectation value of system operators ${\cal O}_A(t) = \tr [\rho_A(t) O_A] = \tr [\rho^{(N=1)}(t) O_A]$ for system operators $O_A$ that conserve the particle number; the latter equality holds because of the vanishing contribution from the vacuum component $\rho^{(N=0)}(t)$ of $\rho_A(t)$. 
	Figure \ref{fig2} shows such results for $O_A=a_1^\dag a_1$ (curves labeled $n_b$) and compares them with the prediction of non-Hermitian dynamics under the operator of Eq. (\ref{LindbladHamiltonian}) with $\Gamma_s=2\frac{C_s^2}{B_s}$  (curve labeled $n_{nH})$. In all cases, the initial state is chosen to be  $a_1^\dag|0\rangle$ and once again we set $C_2=3C_1=3C$, $B_2=2B_1=2B$; similar results are found for any other non-unity ratios. Figure \ref{fig2} demonstrates that as the singular limit is approached with $\frac{C^2}{B}=0.5$ fixed and both $B$ and $C$ increasing, the agreement between the exact results (dashed lines) and the non-Hermitian approximation (solid line) improves steadily.

	For our simple system+bath model, it is straightforward to check that the right eigenvectors for $H_{nh}$ of Eq. (\ref{LindbladHamiltonian}) become identical if
	\begin{equation}
		\label{cd}
		2 A=\frac{1}{2}|\Gamma_1-\Gamma_2|=\left| \frac{C_1^2}{B_1}-\frac{C_2^2}{B_2}\right| .
	\end{equation}
	The semi-infinite chains act as baths when Eq. (\ref{bound}) is satisfied, {\it i.e.} both $B_s/ A$ ratios are larger than a number of order unity.  Thus, Eq. (\ref{cd}) confirms that an exceptional point cannot occur when both couplings are weak. For example,  for our parametrization $C_2=3C_1=3C$, $B_2=2B_1=2B$ we find that Eq. \eqref{cd}  holds if  $3.5 c_A^2=2b_A$. Combining Eqs. \eqref{cd} and \eqref{bound}, we find  that an exceptional point is possible for $b_A>0.733$ and $c_A>0.647$; these can be reached in the singular limit but are well outside the weak coupling limit $c_A \rightarrow 0$.

	One can analyze the time scales required for exceptional points in this 2-site model. In particular, Eq. (\ref{cd}) can be interpreted in terms of three characteristic timescales: $\tau_A \sim \frac{1}{A}$, $\tau_B \sim \frac{1}{B}$ and $\tau_R \sim \frac{1}{\Gamma}\sim \frac{B}{C^2}$ associated with the intrinsic evolution of the system, of the bath, and the relaxation time of the system because of the coupling to the bath, respectively. (For simplicity, we assume that if there are multiple baths, their energy scales and their couplings are proportional to one another). Lindbladian dynamics implies a markovian bath satisfying  $\tau_B \ll \tau_R$; this condition is obeyed both in the weak coupling and in the singular coupling limit. Eq. (\ref{cd}) suggests that an exceptional point can be found if additionally  $\tau_A \sim \tau_R$, {\it i.e.} when the system timescale and the relaxation time are comparable.

	\section{Conclusions} 
	\label{Conclusions}
	We have proven that for Lindblad evolution with only decay terms, the dynamics in the highest particle subspace can be described by a non-Hermitian Hamiltonian. 
	
	We then calculated the exact dynamics of a simple decay model and investigated when it is well approximated by non-Hermitian dynamics in the highest particle subspace considered. Rather surprisingly, we find that this is only true in the weak- and singular-coupling limits where Lindbladian dynamics is already known to be valid. In our opinion, the fact that an accurate non-Hermitian description is so limited even for such a simple system, raises serious questions about how general such descriptions might be for more complicated systems. 
	
	On the other hand, this finding confirms that for these decay systems, the Lindbladian in these known limits has only decay terms, consistent with expectations.  It also means that the evolution of decay models  cannot be accurately described by a Lindbladian with only decay terms anywhere else in the parameter space.

	Finally, we showed that for these bath models with a non-degenerate system Hamiltonian $H_A$, exceptional points cannot occur in the weak coupling limit. This has clear implications for the design of experiments that aim to identify such exceptional points.

	{\it Acknowledgements:} We thank Man-Yat Chu and Riley Duggan for useful comments and suggestions. We acknowledge support from the Max Planck-UBC-UTokyo Center for Quantum Materials and Canada First Research Excellence Fund (CFREF) Quantum Materials and Future Technologies Program of the Stewart Blusson Quantum Matter Institute (SBQMI), and the Natural Sciences and Engineering Research Council of Canada (NSERC).
	
	\appendix 
	\section{Proof of claim R1}
	\label{R1appendix}
	Here we consider a many-body system Hamiltonian of block diagonal form
	\begin{equation}
		\label{Hbasis}
		H=H^{(0)} \oplus H^{(1)} \oplus H^{(2)} \oplus \dots \oplus H^{(N)}.
	\end{equation}
	where each block $H^{(n)}$ represents the $n$ particle Hilbert space. We are interested in density matrices of a similar form
	\begin{equation}
		\label{rho}
		\rho=\rho^{(0)} \oplus \rho^{(1)} \oplus \rho^{(2)} \oplus \dots \oplus \rho^{(N)}.
	\end{equation}
	These assumptions imply that both $H$ and $\rho$ commute with particle number operator as $[H,\hat{N}]=0$ and $[\rho,\hat{N}]=0$.
	
	The Lindblad equation is given by
	\begin{equation}
		\dot{\rho}=-\im [H,\rho] + \sum_j \Gamma_j \left( L_j \rho L_j^\dag -\frac{1}{2} \left\{ L_j^\dag L_j , \rho \right\} \right).
	\end{equation}
	Since we are strictly interested in decay, we will consider the Lindblad operators $L_j$ as annihilation operators. Therefore, they have the form
	\begin{equation}
		\label{Lj}
		L_j=\begin{pmatrix}
			0 & L_j^{(1)} & 0 & \dots & 0 & 0 \\
			0 & 0 & L_j^{(2)} & \dots &  0& 0 \\
			0 & 0 & 0 & \dots & 0 & 0 \\
			\vdots & \vdots & \vdots & \ddots & \vdots & \vdots \\
			0 & 0 & 0 & \dots & 0 & L_j^{(N)} \\
			0 & 0 & 0 & \dots & 0 & 0
		\end{pmatrix}
	\end{equation}
	where the $L_j^{(n)}$ blocks are matrices. With these assumptions on H in eq. \eqref{Hbasis} and $L_j$ in eq. \eqref{Lj}, we can say that if  $\rho$ is initially of the form \eqref{rho}, then it will stay in this form at all times.
	
	Now we analyze how each term of the Lindblad master equation acts on $\rho$. Firstly,
	\small
	\begin{eqnarray}
		[H,\rho] = \hspace{4cm} && \hspace{1cm} \nonumber \\ =\left( [H^{(0)},\rho^{(0)}] \right) \oplus \left( [H^{(1)},\rho^{(1)}] \right) &\oplus& \left( [H^{(2)},\rho^{(2)}] \right) \oplus \dots \nonumber \\ &\dots& \oplus \left( [H^{(N)},\rho^{(N)}] \right). \nonumber \\
	\end{eqnarray}
	\normalsize
	Secondly,
	\small
	\begin{eqnarray}
		\footnotesize
		\nonumber \\
		\left\{ L_j^\dag L_j , \rho \right\} = \hspace{2cm} && \hspace{4cm}\nonumber \\ =
		0 \oplus \left\{ {L_j^{(1)}}^\dag L_j^{(1)} , \rho^1 \right\} &\oplus& \left\{ {L_j^{(2)}}^\dag L_j^{(2)} , \rho^{(2)} \right\} \oplus \dots \nonumber \\ &\dots& \oplus \left\{ {L_j^{(N)}}^\dag L_j^{(N)} , \rho^{(N)} \right\}. \nonumber \\
	\end{eqnarray}
	\normalsize
	We also have
	\small
	\begin{eqnarray}
		L_j \rho L_j^\dag = \hspace{2.8cm} && \hspace{2cm}\nonumber \\ =  (L_j^{(1)} \rho^{(1)} {L_j^{(1)}}^\dag) \oplus (L_j^{(2)} \rho^{(2)} &{L_j^{(2)}}^\dag)& \oplus (L_j^{(3)} \rho^{(3)} {L_j^{(3)}}^\dag) \oplus \dots \nonumber \\ &\dots& \oplus (L_j^{(N)} \rho^{(N)} {L_j^{(N)}}^\dag) \oplus 0. \nonumber \\
	\end{eqnarray}
	\normalsize
	Now we assume that we start with $N$ particles in the system. Therefore $\rho^{(n)}=0$ for $n>N$. Then we have the result for the $\rho^N$
	\begin{equation}
		\dot{\rho}^{(N)}=-\im [H_{\textnormal{eff}},\rho^{(N)}] = -\im \left( H_{\textnormal{eff}} \rho^{(N)} - \rho^{(N)} H_{\textnormal{eff}}^\dag \right)
	\end{equation}
	with the effective Hamiltonian
	\begin{equation}
		\label{Heff}
		H_{\textnormal{eff}}= H^{(N)} -\frac{\im}{2}\sum_j \Gamma_j {L_j^{(N)}}^\dag L_j^{(N)}.
	\end{equation}
	
	\section{Green's functions}
	\label{GreensAppendix}
	Let $H$ be the total Hamiltonian for a subsystem $A$ coupled to a bath model $B$,  let $|a_r\rangle$ be single-particle states in the subsystem $A$, while $|a_r,0_B\rangle$ describes the subsystem A in $|a_r\rangle$ with the bath empty. Furthermore, let $\hat{G}(z)=(z-\hat{H})^{-1}$ and $G_{r,s}(z)=\langle a_r, 0_B| \hat{G}(z) |a_s,0_B\rangle$, where $z=\omega + i\eta$, $\eta \rightarrow 0$ is a small broadening and we set $\hbar=1$. Then the time-dependent retarded Green's function is defined as 
	\begin{eqnarray}
		\label{timeGreen}
		g_{r,s}(t)&=&-\im\Theta(t) \langle a_r,0_B|\exp(-\im H t)| a_s , 0_B\rangle \nonumber \\ &=&\frac{1}{2\pi} \int_{-\infty}^{\infty} d\omega \ e^{-i\omega t} G_{r,s}(z).
	\end{eqnarray}
	where $\Theta(t)$ is the unit step function. 
	
	Next, we define a non-Hermitian Hamiltonian $H_{\text{nH}}$ which acts \textit{only} on subsystem $A$. Then we can define $\hat{\bar{G}}(z)=(z-\hat{H}_{\text{nH}})^{-1}$ and $\bar{G}_{r,s}(z)=\langle a_r| \hat{\bar{G}}(z) |a_s\rangle$. Then similarly
	\begin{eqnarray}
		\label{timeGreennH}
		\bar{g}_{r,s}(t)&=&-\im \Theta(t) \langle a_r|\exp(-\im H_{\text{nH}} t)| a_s \rangle \nonumber \\ &=& \frac{1}{2\pi} \int_{-\infty}^{\infty} d\omega \ e^{-i\omega t} \bar{G}_{r,s}(z).
	\end{eqnarray}

	\subsection{Single site Green's function}
\begin{figure*}[t!]
		\large ($a$)  \ \ \includegraphics[width=0.4\linewidth]{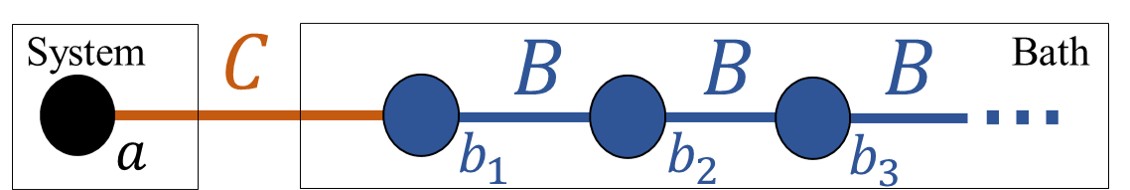}
		\large ($b$) \includegraphics[width=0.35\linewidth]{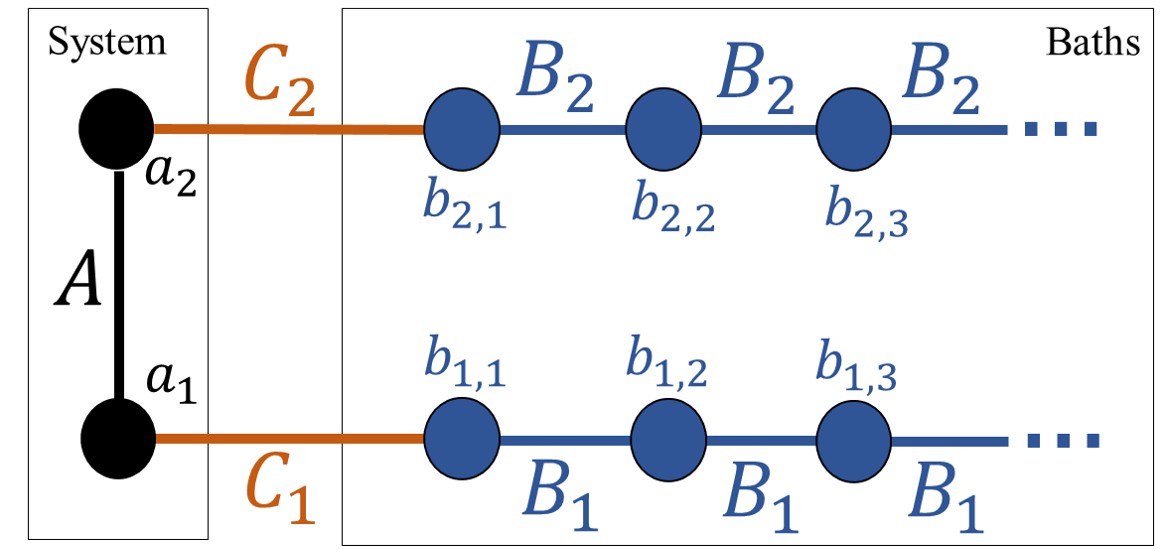} 
		\hfill \hfill \hfill \hfill
		\caption{($a$) Single-site bath model. ($b$) Two-site bath model considered in the main article.}
		\label{fig:supp}
	\end{figure*}

	We consider a single site-Hamiltonian with $H_A=0$, $H_B = B \sum_{j=1}^\infty \left( b_{j}^\dagger b_{j+1}+b_{j+1}^\dag b_{j}  \right)$ and $H_C = C \left( a^\dag b_{1} + b_{1}^\dag a \right)$ shown in Fig. \ref{fig:supp}$a$. The Green's function $G(z)=\langle 0| a \hat{G}(z) a^\dag|0\rangle$ is calculated, as in Ref. \cite{zhu2018excitonic}, to be
	\begin{equation}
		G(z)=\frac{1}{z-\frac{2C^2}{z+\sqrt{z-2B}\sqrt{z+2B}}}.
		\label{singleGreens}
	\end{equation}
	Here we choose the parameters to avoid zeros in the denominator. This is to satisfy a valid-bath condition, which allows the particle to move from the subsystem A into the bath. The condition is
	\begin{equation}
		\label{poleCondition}
		C<\sqrt{2}B.
	\end{equation}
	Define the parameter $\gamma=\frac{C^2}{B}$. Then note that $G(z)$ is bounded
	\begin{equation}
		|G(z)| \leq \max[\frac{1}{\gamma},\frac{1}{2B-\gamma}].
		\label{singleGreenMax}
	\end{equation}
	The difference $2B-\gamma$ is always positive due to eq. \eqref{poleCondition}.
	
	We consider the non-Hermitian Hamiltonian $H_{\text{nH}}=-\im \gamma a^\dag a$ and $\bar{G}_\gamma(z)=\langle a | (z-H_{\text{nH}})^{-1} | a \rangle$. This results in 
	\begin{equation}
		\bar{G}_\gamma(z)=\frac{1}{z+\im \gamma}.
		\label{singleGreensNH}
	\end{equation}

	\subsection{Two Site Green's Function}
	We consider the Hamiltonian of the main article with $H_A = \sum_{r,s}h_{r,s} a_r^\dag a_s =Aa_1^\dag a_2 + A^*a_2^\dag a_1$, $H_B = \sum_{s=1}^2 B_s \sum_{j=1}^\infty \left( b_{s,j}^\dagger b_{s,j+1}+b_{s,j+1}^\dag b_{s,j}  \right)$ and $H_C = \sum_{s=1}^2 C_s \left( a_s^\dag b_{s,1} + b_{s,1}^\dag a_s \right)$. 
	
	The $G_{r,s}(z)$ functions are determined by the coupled equations: 
	\begin{eqnarray}
		\left( z - \frac{2 C_s^2}{z+\sqrt{z-2 B_s}\sqrt{z+2 B_s}} \right) G_{r,s}(z) -\sum_q  h_{q,s} G_{r,q}(z) \nonumber \\
		= \delta_{r,s} \nonumber \\
	\end{eqnarray} \normalsize
	Let 
	\begin{equation}
		S_s(z)=z - \frac{2 C_s^2}{z+\sqrt{z-2 B_s}\sqrt{z+2 B_s}}.
	\end{equation}
	The system of equations for the Greens functions is then
	\begin{eqnarray}
		S_1(z) G_{1,1}(z) - A^* G_{1,2}(z)=1 \ , \ 
		S_2(z) G_{1,2}(z)=A G_{1,1(z)} \nonumber \\
		S_2(z) G_{2,2}(z)-A G_{2,1}(z)=1 \ , \ S_1 (z)G_{2,1}(z)=A^* G_{2,2}(z) \nonumber \\
	\end{eqnarray}
	The solutions are
	\begin{eqnarray}
		\label{GreensTwoSite}
		G_{1,1}(z)=\frac{S_2(z)}{S_1(z) S_2(z)-|A|^2} \ , \ G_{1,2}(z)=\frac{A}{S_1(z) S_2(z)-|A|^2} \nonumber \\
		G_{2,2}(z)=\frac{S_1(z)}{S_1(z) S_2(z)-|A|^2} \ , \ G_{2,1}(z) = \frac{A^*}{S_1(z) S_2(z)-|A|^2}. \nonumber \\
	\end{eqnarray}
	Now we define the Green's functions
	\begin{eqnarray}
		\label{GreensBar}
		\bar{G}_{1,1}(z)=\frac{\bar{S}_2(z)}{\bar{S}_1(z) \bar{S}_2(z)-|A|^2} \ , \ \bar{G}_{1,2}(z)=\frac{A}{\bar{S}_1(z) \bar{S}_2(z)-|A|^2} \nonumber \\
		\bar{G}_{2,2}(z)=\frac{S_1(z)}{\bar{S}_1(z) \bar{S}_2(z)-|A|^2} \ , \ \bar{G}_{2,1}(z) = \frac{A^*}{\bar{S}_1 (z)\bar{S}_2(z)-|A|^2} \nonumber \\
	\end{eqnarray}
	with
	\begin{equation}
		\label{SBar}
		\bar{S}_s(z)=z+\im \gamma_s
	\end{equation}
	These Green's functions $\bar{G}_{r,s}(z)$ are obtained directly from $\bar{G}_{r,s}(z)=\langle a_r |(z-H_{\text{nH}})^{-1}|a_s \rangle$ where the non-Hermitian Hamiltonian $H_{\text{nH}}$ is defined by 
	\begin{equation}
		H_{\text{nH}}=-\im \gamma_1 a_1^\dag a_1 -\im \gamma_2 a_2^\dag a_2 + A(a_1^\dag a_2+a_2^\dag a_1).
	\end{equation}
	and $\gamma_s = \frac{C_s^2}{B_s}$.
	
	\subsubsection{Decay region}
	The requirement for the system to be in the decay region is that the Green's functions do not have zeros in the denominator. That is, 
	\begin{equation}
		S_1(z) S_2 (z)\neq |A|^2
		\label{nopoles}
	\end{equation}
	for all $z \in \mathbb{R}$. 
	
	Now consider our main example with  $C_1=C$, $C_2=3C$, $B_1=B$, $B_2=2B$. Since $A$ could be zero, we expect the single site conditions $C \leq \sqrt{2}B$ and $(3C) \leq \sqrt{2} (2B)$ to be required for decay. In general, the condition for decay will be more complicated in the coupled system. 
	
	The product  $S_1(z) S_2(z)$ is complex in the region $|z|<4B$. Due to these imaginary parts, there cannot be poles in this region. Now consider $4B<|z|$. In this region $S_1(z)$ an $S_2(z)$ are monotonic increasing functions. Thus if $S_1 (z)S_2(z) > |A|^2$ at $z=4B$, then there will be no zeros in any region. To determine the parameter bounds we check when $S_1(z) S_2(z)=|A|^2$ for $z=4B$. This condition is
	\begin{eqnarray}
		\label{zero}
		(4B-\frac{2 C^2}{4B+\sqrt{2B}\sqrt{6B}})(4B-\frac{18 C^2}{4B})=|A|^2.
	\end{eqnarray}
	Define $b_A=B/A$, $c_A=C/A$, $x_0=1/(2+\sqrt{3})$ and $y_0=9/2$. Then we can rewrite eq. \eqref{zero} as 
	\begin{equation}
		\label{zero2}
		c_A^2=\frac{2b_A^2(x_0+y_0) \pm \sqrt{4b_A^4(x_0-y_0)^2+x_0 y_0 b_A^2}}{x_0 y_0}
	\end{equation}
	Considering the decay requirement $S_1 S_2 > |A|^2$, we have the condition for no zeros in the denominator as
	\begin{equation}
		\label{mainDecayInequality}
		c_A^2<\frac{2b_A^2(x_0+y_0) - \sqrt{4b_A^4(x_0-y_0)^2+x_0 y_0 B_A^2}}{x_0 y_0}.
	\end{equation}
	This defines the blue curve plotted in Fig. 2 of the main article. This condition implies a number of other inequalities
	\begin{equation}
		\label{AIndependent}
		C < \sqrt{2} B \ , \ (3C) < \sqrt{2} (2B) \ , \ 9 \gamma < 8 B \ , \ A<4B.
	\end{equation}
	
	\subsubsection{Numerical Methods}
	Consider the Green's functions $\bar{G}_{r,s}(z)$ defined in eqs. \eqref{GreensBar} and \eqref{SBar}. In the singular and weak coupling limits, these Green's functions match up with $G_{r,s}(z)$ defined in eq. \eqref{GreensTwoSite}. However, even outside of the two limits, the function $\bar{G}_{r,s}(z)$ can be used to help the numerics converge better.

	The $\bar{G}_{r,s}(z)$ functions have known Fourier transforms when plugged into eq. \eqref{timeGreennH}. That is, we have \small
	\begin{eqnarray}
		\label{barGreenTime} 
		\bar{g}_{1,1}(t)&=&\frac{-\im}{Z_+ - Z_-} \left( e^{-\im Z_+ t}(Z_+ +\im \frac{C_2^2}{B_2}) - e^{-\im Z_- t}(Z_- +\im \frac{C_2^2}{B_2}) \right) \nonumber \\
		\bar{g}_{1,2}(t)&=&\frac{-\im A}{Z_+ - Z_-} \left( e^{-\im Z_+ t} - e^{-\im Z_- t} \right) \nonumber \\
		\bar{g}_{2,1}(t)&=&\frac{-\im A^*}{Z_+ - Z_-} \left( e^{-\im Z_+ t} - e^{-\im Z_- t} \right) \nonumber \\
		\bar{g}_{2,2}(t)&=&\frac{-\im}{Z_+ - Z_-} \left( e^{-\im Z_+ t}(Z_+ +\im \frac{C_1^2}{B_1}) - e^{-\im Z_- t}(Z_- +\im \frac{C_1^2}{B_1}) \right) \nonumber \\
	\end{eqnarray} \normalsize
	with
	\begin{eqnarray}
		Z_\pm=\frac{ -\im (\gamma_1+\gamma_2) \pm \sqrt{4|A|^2-(\gamma_1-\gamma_2)^2} }{2}. \nonumber \\
	\end{eqnarray}
	
	There are numerical challenges if $Z_+=Z_-$. However, we have checked numerically that Fig. 2 of the main article does not have this exact condition for any of the discretely evaluated grid points. 
	
	Now we numerically take the Fourier transform of
	\begin{equation}
		\Delta G_{r,s}(z)=G_{r,s}(z)-\bar{G}_{r,s}(z)
	\end{equation}
	with the limits of the energy integration as $-D$ to $D$ with $D=800$. We call the Fourier transform of $\Delta G_{r,s}(z)$ as $\Delta g_{r,s}(t)$. Then we take in our analysis the Green's functions
	\begin{equation}
		g_{r,s}(t)=\bar{g}_{r,s}(t)+\Delta g_{r,s}(t)
	\end{equation}
	That is, the $g_{r,s}(t)$ we use in the main text is the sum of the analytical $\bar{g}_{r,s}(t)$ of eq. \eqref{barGreenTime} plus the numerically calculated $\Delta g_{r,s}(t)$. This process is numerically well behaved.
	
	\section{Claim R3}
	\label{R3Appendix}
	R3 can be broken down into two separate parts. The first is that exceptional points cannot occur in the weak coupling limit if the spectrum of $H_A$ is non-degenerate. For this, we present a general perturbative argument.

	The second statement is that exceptional points can occur in the singular coupling limit. Since we say that it \textit{can} occur, we only need to show one example of it occurring. To prove this, we show that the singular coupling dynamics exactly matches Lindblad dynamics in the model of the main paper. Then since there are exceptional points in the Lindblad and non-Hermitian descriptions, there are exceptional points in the singular-coupling limit dynamics. 
	
	\subsection{Exceptional points are not in the weak limit}
	
	In this section, we show that for any dynamical evolution in the weak coupling limit and non-degenerate $H_A$, there is corresponding non-Hermitian evolution which is arbitrarily close for all $t>0$. Importantly, the corresponding non-Hermitian Hamiltonian is diagonal in an orthogonal basis. Therefore, the number of eigenstates is equal to the size $D$ of the subsystem $A$. This means that the system is not at an exceptional point since an exceptional point would correspond to fewer than $D$ eigenstates.

	The $G_{r,s}(z)$ functions for a general system with a bath-chain connected to each states are determined by the coupled equations: 
	\begin{eqnarray}
		\left( z - \frac{2 C_s^2}{z+\sqrt{z-2 B_s}\sqrt{z+2 B_s}} \right) G_{r,s}(z) -\sum_q  A_{q,s} G_{r,q}(z) \nonumber \\
		= \delta_{r,s} \nonumber \\
		\label{generalGreens}
	\end{eqnarray}
	To have a uniform limit, we let $C_s^2=\Delta \zeta_s^2$ where $\zeta_s$ are constants and $\Delta \rightarrow 0$ in the weak coupling limit. Define a matrix $\Gamma_{r,s}(z)=\frac{2 \delta_{r,s} \zeta_s^2}{z+\sqrt{z-2 B_s}\sqrt{z+2 B_s}}$. Then we can rewrite equation \eqref{generalGreens} in matrix form as
	\begin{equation}
		G(\Delta, z)=(z \bold{1} -H_A-\Delta \Gamma)^{-1}
	\end{equation}
	where $\bold{1}$ is the identity matrix. Since we are dealing with only systems in the decay region, we require that $G(\Delta, z)$ does not have poles on the real line. More specifically, they are only in the lower region of the complex plane. 
	
	Now we want to expand the eigenvalues and eigenvectors of both $G(\Delta,z)$ and $H=H_A+\Delta \Gamma$. Since $H$ is perturbed around a Hermitian matrix $H_A$ with orthogonal eigenvectors, the eigenpairs are holomorphic as a function of $\Delta$. That is, we can use non-degenerate perturbation theory to expand the eigenpairs of $H$ as 
	\begin{eqnarray}
		E_j=E_j^{(0)}+\Delta E_j^{(1)}+\Delta^2 E_j^{(2)}+\dots \\
		|E_j\rangle=|E_j^{(0)}\rangle+\Delta |E_j^{(1)}\rangle+\Delta^2 |E_j^{(2)} \rangle+\dots
	\end{eqnarray}
	where $E_j^{(0)}$ and $|E_j^{(0)}\rangle$ are eigenpairs of $H_A$. 
	
	The Green's function $G(\Delta,z)$ is known as the resolvent of $H_A+\Delta \Gamma$. Due to the decay condition, the eigenvalues of $H_A+\Delta \Gamma$ are never equal to $z$, which is integrated over the real line. Therefore, $G(\Delta,z)$ is holomorphic in $\Delta$ as well \cite{kato2013perturbation}. That is, $G(\Delta, z)$ has eigenpairs  
	\begin{eqnarray}
		G_j=G_j^{(0)}+\Delta G_j^{(1)}+\Delta^2 G_j^{(2)}+\dots \\
		|G_j\rangle=|G_j^{(0)}\rangle+\Delta |G_j^{(1)}\rangle+\Delta^2 |G_j^{(2)} \rangle+\dots
	\end{eqnarray}
	where $G_j^{(0)}$ and $|G_j^{(0)}\rangle$ are eigenpairs of $(z-H_A)^{-1}$. 
	
	We also note a few important relationships. That is, $|G_j\rangle=|E_j\rangle$ is the same expansion. Furthermore, the states $|E_j\rangle$ and the eigenvalues $E_j$ are not dependent on the $z$ variable. We also have that $G_j=(z-E_j)^{-1}$.
	
	Then we have
	\begin{eqnarray}
		\langle E_j | \hat{G}(\Delta,z) | E_i \rangle=G_i \langle E_j | E_i \rangle=(z-E_i)^{-1} \langle E_j | E_i \rangle \nonumber \\
	\end{eqnarray}
	
	Let O$(\Delta)$ refer to the order in $\Delta$. Then
	\begin{eqnarray}
		\langle E_j | \hat{g}(t) &|E_i\rangle&=\frac{1}{2\pi} \int_{-\infty}^{\infty} dz \ e^{-i z t} \langle E_j | \hat{G}(z) |E_i\rangle = \nonumber \\
		&=&\frac{\langle E_j |E_i\rangle}{2\pi} \int_{-\infty}^{\infty} dz \ e^{-i z t}  (z-E_i)^{-1} \nonumber \\
		&=&\langle E_j |E_i\rangle e^{-\text{i} z E_i t} \nonumber \\
		&=&\langle E_j^{(0)} |E_i^{(0)}\rangle e^{-\text{i} z E_i t} + \text{O}(\Delta) \nonumber \\    
		&=&\delta_{i,j} \ e^{-\text{i} z E_i t} + \text{O}(\Delta) \nonumber \\    
		&=&\delta_{i,j} \ e^{-\text{i} zt E_i^{(0)} } e^{-\text{i} zt (\Delta E_i^{(1)}+\Delta^2 E_i^{(2)} + \dots )} + \text{O}(\Delta) \nonumber \\
	\end{eqnarray}
	
	Using this relation, we can look at $g(t)$ in the orthogonal basis of eigenstates of $H_A$. We have
	\begin{eqnarray}
		\langle E_j^{(0)} |&& \hat{g}(t) |E_i^{(0)}\rangle=\langle E_j | \hat{g}(t) |E_i\rangle+\text{O}(\Delta) \nonumber \\
		&&=\delta_{i,j} \ e^{-\text{i} zt E_i^{(0)} } e^{-\text{i} zt (\Delta E_i^{(1)}+\Delta^2 E_i^{(2)} + \dots )} + \text{O}(\Delta) \nonumber \\
	\end{eqnarray}
	
	Now consider a non-Hermitian matrix $H_{\text{nh}}$ defined by
	\begin{eqnarray}
		\langle E_j^{(0)} |&& H_{\text{nh}} |E_i^{(0)}\rangle=\delta_{i,j} E_i \nonumber \\
		&&=\delta_{i,j} E_i^{(0)}+\delta_{i,j} (\Delta E_i^{(1)} +\Delta^2 E_i^{(2)}+\dots)
	\end{eqnarray}
	Then define Green's function $\hat{\bar{G}}(z)=(z-H_{\text{nh}})^{-1}$. Then the time evolution $\hat{\bar{g}}(t)$ is  given by
	\begin{eqnarray}
		\langle E_j^{(0)} | \hat{\bar{g}}(t) &|E_i^{(0)}\rangle& =\delta_{i,j} \ e^{-\text{i} z E_i t} \nonumber \\    
		&=&\delta_{i,j} \ e^{-\text{i} zt E_i^{(0)} } e^{-\text{i} zt (\Delta E_i^{(1)}+\Delta^2 E_i^{(2)} + \dots )}  \nonumber \\
	\end{eqnarray}
	
	This then implies that the dynamics of $g(t)$ and $\bar{g}(t)$ are arbitrarily close to each other. That is, 
	\begin{eqnarray}
		|\langle E_j^{(0)} | \hat{g}(t) |E_i^{(0)}\rangle-\langle E_j^{(0)} | \hat{\bar{g}}(t) |E_i^{(0)}\rangle| \approx \text{O}(\Delta) \rightarrow 0 \nonumber \\
	\end{eqnarray}
	as $\Delta \rightarrow 0$.
	
	Thus the dynamics of the weak coupling limit become exponentially close to that of the non-Hermitian dynamics of $H_{\text{nh}}$ for any $t>0$. $H_{\text{nh}}$ is diagonal in an orthonormal basis $|E_i^{(0)} \rangle$ and therefore is not at an exceptional point.

	\subsection{Exceptional points can occur in the singular coupling limit}
	Here we show that the singular coupling dynamics exactly matches Lindblad dynamics in the model of the main paper. This then shows a part of R3: since there are exceptional points in the Lindblad/non-Hermitian description, there are exceptional points in the singular-coupling limit dynamics.

	The question now is whether there exists a non-Hermitian operator $H_{\text{nh}}$ such that $g_{r,s}(t) \rightarrow \bar{g}_{r,s}(t)$ in the singular limit. We will show that this is true and that one can select any fixed Lindbladian decay parameter $2\Gamma_s=\frac{C_s^2}{B_s} \equiv \gamma_s$. Specifically, we will show that $g_{r,s}(t)$ converges uniformly to $\bar{g}_{r,s}(t)$ in the singular coupling limit. That is, we show that $|\Delta g(t)| = |g(t)-\bar{g}(t)|$ can be made arbitrarily small for all values of $t$. Since any $\Gamma_s$ can be specified, exceptional points can always be reached. 
	
	Below we assume that $z$ is real, since we are in the decay region without poles on the real line. Then, we  define $\Delta G(z) = G(z)-\bar{G}(z)$. Then 
	\begin{eqnarray}
		\label{mainInequality}
		&2 \pi |\Delta g(t)|& = 2\pi  | \Re \Delta g(t) + \im \Im \Delta g(t) | \nonumber \\
		&\leq&  2 \pi | \Re \Delta g(t) | + 2 \pi |\Im \Delta g (t)| \nonumber \\
		&\leq& 4 \pi \int_{-\infty}^{\infty} dz | \Re \Delta G(z) |+4 \pi \int_{-\infty}^{\infty} dz | \Im \Delta G(z) |. \nonumber \\
	\end{eqnarray}
	Therefore, if we can show that $\int_{-\infty}^{\infty} dz | \Re \Delta G(z) |$ and $\int_{-\infty}^{\infty} dz | \Im \Delta G (z)|$ can be made arbitrarily small, this is sufficient to ensure uniform convergence. 
	
	First we show this for a single site system Green's functions, where $G(z)$ and and $\bar{G}(z)$ are defined by eqs. \eqref{singleGreens} and \eqref{singleGreensNH} respectively. 
	Then we have that 
	\begin{eqnarray}
		\Re \Delta G(z) =\begin{cases}
			\frac{\gamma}{2B} \frac{z(z^2-\gamma^2)}{(z^2+\gamma^2)(z^2(1-\frac{\gamma}{B})+\gamma^2)}  \text{\ \ \ \ \ \ \ \ \ \ \ if  \ } 0 \leq z \leq 2B \\ \\
			\frac{\gamma}{2B} \frac{z(z-\sqrt{z^2-4B^2})+2B\gamma}{(z^2+\gamma^2)(z(1-\frac{\gamma}{2B})+\frac{\gamma}{2B}\sqrt{z^2-4B^2})} \text{\ \ \ if \ } 2B < z
		\end{cases} \nonumber \\
	\end{eqnarray} 
	$\Re \Delta G(z)$ is an odd function. Thus, $\Re \Delta G(z)=\Re \Delta G(-z)$ defines the function in the interval $z<0$. This implies then that $| \Re \Delta G |$ is even. Thus, we need to show that the positive part of the integral can be made arbitrarily small in the singular limit. 
	
	Note that \footnotesize
	\begin{eqnarray} 
		|\frac{\gamma}{2B} \frac{z(z^2-\gamma^2)}{(z^2+\gamma^2)(z^2(1-\frac{\gamma}{B})+\gamma^2)}| \leq \frac{\gamma}{2B} \frac{z(z^2+\gamma^2)}{(z^2(1-\frac{\gamma}{B})+\gamma^2)^2}. \nonumber \\
	\end{eqnarray} \normalsize
	
	Thus \footnotesize
	\begin{eqnarray}
		\int_{0}^{2B} dz | \Re \Delta G (z)| \leq \int_{0}^{2B} dz \frac{\gamma}{2B} \frac{z(z^2+\gamma^2)}{(z^2(1-\frac{\gamma}{B})+\gamma^2)^2} \nonumber \\
		=\frac{\gamma}{4B(1-\frac{\gamma}{B})^2} \left( \frac{\gamma^3}{B(\gamma^2+(1-\frac{\gamma}{B})z^2)} +\log(\gamma^2+(1-\frac{\gamma}{B})z) \right) \Bigg{|}_0^{2B} \nonumber \\
	\end{eqnarray}
	\normalsize
	which scales as $\frac{\log B}{B}\rightarrow 0$ when $B\rightarrow \infty$.
	
	Now we consider at $\int_{2B}^{\infty} dz \ | \Re \Delta G(z)|$. We have \footnotesize
	\begin{eqnarray}
		\label{ReGSecond}
		&&\int_{2B}^{\infty} dz \ | \Re \Delta G (z)| < \frac{\gamma}{2B} \int_{2B}^{\infty} dz \ \frac{z(z-\sqrt{z^2-4B^2})+2B\gamma}{(z^2+\gamma^2)(2B-\gamma)} \nonumber \\
		&&< \int_{2B}^{\infty} dz \ \left( \frac{\gamma(z-\sqrt{z^2-4B^2})}{2B(2B-\gamma)z}+ \frac{\gamma^2}{(z^2+\gamma^2)(2B-\gamma)} \right). \nonumber \\
		&&=\frac{\gamma}{2B(2B-\gamma)} \left( (z-\sqrt{z^2-4B^2})+2B \tan^{-1} \frac{\sqrt{z^2-4B^2}}{2B} \right)\Bigg{|}_{2B}^{\infty} +  \nonumber \\
		&&+ \left( \frac{\gamma \tan^{-1} (\frac{z}{\gamma})}{2B-\gamma} \right) \Bigg{|}_{2B}^{\infty} \nonumber \\
	\end{eqnarray} 
	\normalsize
	which goes to zero as $\frac{1}{B}$.
	
	Next, we consider $\Im \Delta G(z)$. We have 
	\begin{eqnarray}
		\Im \Delta G(z) =\begin{cases}
			\frac{\gamma}{2B} \frac{(z^2+\gamma^2)(2B-\sqrt{4B^2-z^2})-2\gamma z^2}{(z^2+\gamma^2)(z^2(1-\frac{\gamma}{B})+\gamma^2)}  \text{\ \ \ if  \ } 0 \leq z \leq 2B \\ \\
			\ \ \ \ \ \ \ \ \ \ \ \ \ \ \ \ \ \frac{\gamma}{z^2+\gamma^2} \text{\ \ \ \ \ \ \ \ \ \ \ \ \ \ \ \ \ if \ } 2B < z
		\end{cases} \nonumber \\\nonumber
	\end{eqnarray} 
	$\Im \Delta G(z)$ is an even function. Thus, $\Im \Delta G(z)=\Im \Delta G(-z)$ defines the function in the interval $z<0$. This implies that $| \Im \Delta G |$ is also even. Similarly to the real case, we need to show that 
	\begin{equation}
		\int_{0}^{\infty} dz \ | \Im \Delta G|=\int_{0}^{2B} dz \ | \Im \Delta G| +\int_{2B}^{\infty} dz \ | \Im \Delta G|  
	\end{equation}
	can be made arbitrarily small in the singular coupling limit.
	
	We start by looking at $\int_{0}^{2B} dz \ | \Im \Delta G|$. In the singular limit, we can safely say that $\frac{\gamma}{B}<0.5$. Furthermore, one can show that $2B-\sqrt{4B^2-z^2} \leq \frac{z^2}{4B}$ in this interval. Then \scriptsize
	\begin{eqnarray}
		\label{ImBound}
		&&\int_{0}^{2B} dz \ | \Im \Delta G(z)|  \nonumber \\ 
		&&< \int_{0}^{2B} dz \ \frac{\gamma}{2B} \left( \frac{(2B-\sqrt{4B^2-z^2})}{ z^2 (1-\frac{\gamma}{B}) + \gamma^2} + \frac{2 \gamma z^2}{( z^2 (1-\frac{\gamma}{B})+ \gamma^2)^2} \right) \nonumber \\
		&&< \int_{0}^{2B} dz \ \frac{\gamma}{2B} \left( \frac{z^2}{4B( z^2 (1-\frac{\gamma}{B})+ \gamma^2)} + \frac{2 \gamma z^2}{(z^2 (1-\frac{\gamma}{B}) + \gamma^2)^2} \right) \nonumber \\
		&&=\frac{\gamma}{8B^2} \left( \frac{z}{1-\frac{\gamma}{B}} - \frac{\gamma}{(1-\frac{\gamma}{B})^{3/2}} \tan^{-1}(\frac{z}{\gamma}\sqrt{1-\frac{\gamma}{B}}) \right) \Bigg{|}_{0}^{2B} + \nonumber \\
		&&+\frac{\gamma^2}{B} \left( \frac{-z}{2(1-\frac{\gamma}{B})((1-\frac{\gamma}{B})z^2+\gamma^2)} + \frac{\tan^{-1}(\frac{z}{\gamma}\sqrt{1-\frac{\gamma}{B}})}{2\gamma(1-\frac{\gamma}{B})^{3/2}}  \right) \Bigg{|}_{0}^{2B} \nonumber \\
	\end{eqnarray}
	\normalsize
	The right hand side of this inequality scales as $\frac{1}{B}$, which goes to zero as $B$ increases.
	
	Next we look at $\int_{2B}^{\infty} dz \ | \Im \Delta G(z)|$. We can solve this analytically. We have 
	\begin{eqnarray}
		\int_{2B}^{\infty} dz \ | \Im \Delta G(z)|=\int_{2B}^{\infty} dz \ \frac{\gamma}{z^2+\gamma^2} \nonumber \\
		=\tan^{-1} (\frac{z}{\gamma}) \Big|_{2B}^\infty=\frac{\pi}{2}-\tan^{-1} (\frac{2B}{\gamma})
	\end{eqnarray}
	which goes to zero as $B$ increases.
	
	Thus, using these results and the inequality \eqref{mainInequality}, we have shown that $|\Delta g(t)|$ can be made arbitrarily small in the singular coupling limit. 
	
	For the two site system, we show the result for $\Delta G_{1,1}(z)=G_{1,1}(z)-\bar{G}_{1,1}(z)$; the proof for other matrix elements is similar. We first define $G_s(z)=(z - \frac{2 C_s^2}{z+\sqrt{z-2 B_s}\sqrt{z+2 B_s}})^{-1}$ for $s=1,2$. The $G_s(z)$ elements are the Green's functions if the sites were independent ($A=0$). We have 
	\begin{eqnarray}
		\Delta G_{1,1}(z)&=& \frac{G_1(z)}{1-|A|^2 G_1(z) G_2(z)}-\frac{\bar{G}_1(z)}{1-|A|^2 \bar{G}_1(z) \bar{G}_2(z)} \nonumber \\ &=&
		\frac{\Delta G_1-|A|^2 G_1(z) \bar{G}_1(z) \Delta G_2}{(1-|A|^2 G_1(z) G_2(z))(1-|A|^2 \bar{G}_1(z) \bar{G}_2(z))}. \nonumber \\
	\end{eqnarray} 
	Using this, we will show that 
	\begin{eqnarray}
		\Delta G_{1,1}(z) \leq (\text{const}_1) \Delta G_1 + (\text{const}_2) \Delta G_2
	\end{eqnarray}
	for bounded constants $\text{const}_1$ and $\text{const}_2$. Then since we know that $\int dz \Delta G_1(z)$ and $\int dz \Delta G_2(z)$ are bounded, the prior results finishes the proof. 
	
	Now we use the result eq. \eqref{singleGreenMax}, which in the singular limit we have that $G_s(z) \leq \frac{1}{\gamma_s}$ and $\bar{G}_s(z) \leq \frac{1}{\gamma_s}$. Given this, we need to only show that $(1-|A|^2 G_1(z) G_2(z))^{-1}$ and $(1-|A|^2 \bar{G}_1(z) \bar{G}_2(z))^{-1}$ are bounded.
	
	To take the limits, we define $B$ such that $B_1=B$ and $B_2=\beta B$ and take the limit as $B \rightarrow \infty$. Without loss of generality we can assume $\beta>1$ so that $B_2$ is larger than $B_1$. We also have the decay conditions $1>|A|^2 G_1(z) G_2(z)$ and $1>|A|^2 \bar{G}_1(z) \bar{G}_2(z)$. Then \small
	\begin{eqnarray}
		\frac{1}{|1-|A|^2 \bar{G}_1 \bar{G}_2|^2} 
		=\left(1-\frac{2|A|^2 ( z^2-\gamma_1 \gamma_2-|A|^2/2)}{(z^2-\gamma_1 \gamma_2)^2+z^2 (\gamma_1+\gamma_2)^2} \right)^{-1} \nonumber \\
	\end{eqnarray} \normalsize
	To find an upper bound to the above, we only need to look in the interval where $2 (z^2-\gamma_1 \gamma_2)-|A|^2 \geq 0$, which we will assume to be the case from now. Consider the inequality $\frac{2D_1-D_2}{D_1^2+D_3} \leq \frac{2}{D_2+\sqrt{D_2^2+4D_3}}$ for any positive real numbers $D_1$, $D_2$ and $D_3$, which can be proven by optimizing with respect to $D_1$. Then \footnotesize
	\begin{eqnarray} 
		&&\frac{|A|^2 (2 (z^2-\gamma_1 \gamma_2)-|A|^2)}{(z^2-\gamma_1 \gamma_2)^2+z^2 (\gamma_1+\gamma_2)^2} \leq \frac{|A|^2 (2 (z^2-\gamma_1 \gamma_2)-|A|^2)}{(z^2-\gamma_1 \gamma_2)^2+\gamma_1 \gamma_2 (\gamma_1+\gamma_2)^2} \nonumber \\
		&&\leq \frac{2|A|^2}{|A|^2+\sqrt{|A|^4+4\gamma_1 \gamma_2 (\gamma_1+\gamma_2)^2}} 
	\end{eqnarray} \normalsize
	
	Thus \small
	\begin{eqnarray}
		\frac{1}{|1-|A|^2 \bar{G}_1 \bar{G}_2|^2} \leq 
		\left(1-\frac{2|A|^2}{|A|^2+\sqrt{|A|^4+4\gamma_1 \gamma_2 (\gamma_1+\gamma_2)^2}} \right)^{-1} \nonumber 
	\end{eqnarray} \normalsize
	which is a bounded constant. 
	
	Next we consider $|(1-|A|^2 G_1(z) G_2(z))|^{-2}$. This function is even so we only need to consider $z \geq 0$. 
	
	Consider the interval $0 \leq z \leq 2B_1$. Define $z_1'=z(1-\frac{\gamma_1}{2B_1})$, $\gamma_1'=\gamma_1\sqrt{1-(\frac{z}{2B_1})^2}$, $z_2'=z(1-\frac{\gamma_2}{2B_2})$ and $\gamma_2'=\gamma_2\sqrt{1-(\frac{z}{2B_2})^2}$. Then 
	\begin{eqnarray}
		\label{0toB}
		\frac{1}{|1-|A|^2 G_1 G_2|^2} 
		=\left(1-\frac{|A|^2 (2 (z_1' z_2'-\gamma_1' \gamma_2')-|A|^2)}{(z_1' z_2'-\gamma_1' \gamma_2')^2+(z_1' \gamma_2'+z_2' \gamma_1')^2} \right)^{-1}. \nonumber 
	\end{eqnarray}
	In this interval $0 \leq \gamma_1' \leq \gamma_1$ and $\sqrt{1-\frac{1}{\beta^2}} \gamma_2 \leq \gamma_2' \leq \gamma_2$. The maximum of this function is certainly in the interval for which $z_1' z_2' \geq \gamma_1' \gamma_2'+|A|^2/2$ and we will assume this to be the case from now on (otherwise the maximum is one). This is equivalent to $z^2\geq z_{min}^2 \equiv (\gamma_1' \gamma_2'+|A|^2/2)/( (1-\frac{\gamma_1}{2B_1}) (1-\frac{\gamma_1}{2B_2}) )$ which is a bounded minimum as $( (1-\frac{\gamma_1}{2B_1}) (1-\frac{\gamma_1}{2B_2}) )$ is arbitrarily close to 1 in the singular limit. 
	
	Then we have \small
	\begin{eqnarray}
		&&\frac{|A|^2 (2 (z_1' z_2'-\gamma_1' \gamma_2')-|A|^2)}{(z_1' z_2'-\gamma_1' \gamma_2')^2+(z_1' \gamma_2'+z_2' \gamma_1')^2} \nonumber \\
		&&\leq \frac{|A|^2 (2 (z_1' z_2'-\gamma_1' \gamma_2')-|A|^2)}{(z_1' z_2'-\gamma_1' \gamma_2')^2+z^2 (1-\frac{\gamma_1}{2B_1})^2 \gamma_2'^2} \nonumber \\
		&&\leq \frac{2|A|^2}{|A|^2+\sqrt{|A|^4+4 z^2 (1-\frac{\gamma_1}{2B_1})^2 \gamma_2'^2}} \nonumber \\
		&&\leq \frac{2|A|^2}{|A|^2+\sqrt{|A|^4+4 z_{\min}^2 (1-\frac{\gamma_1}{2B_1})^2 \gamma_2'^2}} \nonumber \\
	\end{eqnarray} \normalsize
	which is a bounded constant.
	
	Now consider the interval $2B_1 \leq z \leq 2B_2$. Let $z_2'=z(1-\frac{\gamma_2}{2B_2})$, $\gamma_2'=\gamma_2 \sqrt{1-(\frac{z}{2B_2})^2}$ and $S_1=z(1-\frac{\gamma_1}{2B_1})+\gamma_1 \sqrt{(\frac{z}{2B_1})^2-1}$. Then $0 \leq \gamma_2' \leq \gamma_2$, $2B_1-\gamma_1 \leq S_2 < \infty$ and $0 < \frac{1}{S_1} \leq \frac{1}{2B_1-\gamma_1}$. Then 
	\begin{eqnarray}
		\frac{1}{|1-|A|^2 G_1 G_2|^2} 
		=\left(1-\frac{|A|^2(|A|^2-2 S_1 z_2')}{S_1^2 (z_2'^2+\gamma_2'^2)} \right)^{-1}. \nonumber \\
	\end{eqnarray}
	We consider $|A|^2-2 S_1 z_2'>0$ (otherwise the maximum of $\frac{1}{|1-|A|^2 G_1 G_2|^2}$ is $1$). Using again the relation $\frac{2D_1-D_2}{D_1^2+D_3} \leq \frac{2}{D_2+\sqrt{D_2^2+4D_3}}$, we have 
	\begin{eqnarray}
		&&\frac{|A|^2(|A|^2-2 S_1 z_2')}{S_1^2 (z_2'^2+\gamma_2'^2)} \leq \frac{2|A|^2}{S_1^2(|A|^2+\sqrt{|A|^2+4\gamma_2'^2})} \nonumber \\
		&&\leq \frac{2|A|^2}{S_1^2} \leq \frac{2|A|^2}{(2B_1-\gamma_1)^2}.
	\end{eqnarray}
	Then 
	\begin{eqnarray}
		\frac{1}{|1-|A|^2 G_1 G_2|^2} 
		\leq \left(1-\frac{2|A|^2}{(2B_1-\gamma_1)^2} \right)^{-1} \nonumber \\
	\end{eqnarray}
	which is arbitrarily close to 1 when $B=B_1\rightarrow \infty$. 
	
	Lastly, consider $2B_2<z$. Let $S_1=z(1-\frac{\gamma_1}{2B_1})+\gamma_1 \sqrt{(\frac{z}{2B_1})^2-1}$ and $S_2=z(1-\frac{\gamma_2}{2B_2})+\gamma_2 \sqrt{(\frac{z}{2B_2})^2-1}$. We then have $(2B_2-\beta \gamma_1)+\gamma_1 \sqrt{\beta^2-1} \leq S_1<\infty$ and $2B_2-\gamma_2\leq S_1<\infty$. So then
	\begin{eqnarray}
		\frac{1}{|1-|A|^2 G_1 G_2|^2} 
		=\left( 1-\frac{|A|^2}{S_1 S_2} \right)^{-2}. \nonumber \\
		\leq \left( 1-\frac{|A|^2}{((2B_2-\beta \gamma_1)+\gamma_1 \sqrt{\beta^2-1}) (2B_2-\gamma_2)} \right)^{-2}
	\end{eqnarray}
	which approaches $1$ from below as $B \rightarrow \infty$.
	
	Thus we have shown that
	\begin{eqnarray}
		\Delta G_{1,1}(z) \leq (\text{const}_1) \Delta G_1 + (\text{const}_2) \Delta G_2
	\end{eqnarray}
	in all intervals. Furthermore, we have shown that the integrals of $\Delta G$ go to zero in the singular limit. 
	\bibliography{bibliography}
\end{document}